\documentclass{svproc}
\usepackage{url}

\usepackage[T1]{fontenc}
\usepackage[latin9]{inputenc}
\usepackage{array}
\usepackage{verbatim}
\usepackage{multirow}
\usepackage{amsbsy}
\usepackage{amssymb}
\usepackage{graphicx}
\usepackage[numbers]{natbib}

\begin{document}
\mainmatter              
\title{Delayed Acceptance Markov Chain Monte Carlo for Robust Bayesian Analysis}
\titlerunning{DA--MCMC for Robust Bayesian Analysis}  
%
\author{Masahiro Tanaka}
%
%
%
\institute{Fukuoka University, 8-19-1, Nanakuma, Jonan,
Fukuoka 814-0180, Japan\\
\email{m.tanaka.tt@fukuoka-u.ac.jp}
}

\maketitle              

\begin{abstract}
This study introduces a computationally efficient algorithm, delayed
acceptance Markov chain Monte Carlo (DA--MCMC), designed to improve
posterior simulation in quasi-Bayesian inference. Quasi-Bayesian methods,
which do not require fully specifying a probabilistic model, are often
computationally expensive owing to the need to evaluate the inverse
and determinant of large covariance matrices. DA--MCMC addresses
this challenge by employing a two-stage process: In the first stage,
proposals are screened using an approximate posterior, whereas a final
acceptance or rejection decision is made in the second stage based
on the exact target posterior. This reduces the need for costly matrix
computations, thereby improving efficiency without sacrificing accuracy.
We demonstrate the effectiveness of DA--MCMC through applications
to both synthetic and real data. The results demonstrate that, although
DA--MCMC slightly reduces the effective sample size per iteration
compared with the standard MCMC, it achieves substantial improvement
in terms of effective sample size per second, approximately doubling
the efficiency. This makes DA--MCMC particularly useful for cases
where posterior simulation is computationally intensive. Thus, the
DA--MCMC algorithm offers a significant advancement in computational
efficiency for quasi-Bayesian inference, making it a valuable tool
for robust Bayesian analysis.
\keywords{delayed acceptance Markov chain Monte Carlo; robust Bayesian
analysis; generalized posterior; Gibbs posterior}
\end{abstract}

\section{Introduction}

In standard Bayesian inference, a probabilistic model is used for
the data-generating process and the corresponding likelihood function,
$p\left(\mathcal{D}|\boldsymbol{\theta}\right)$, where $\mathcal{D}$
denotes a set of observations with a sample size of $N$, $\boldsymbol{\theta}\in\mathbb{R}^{K}$
denotes a $K$-dimensional vector of unknown parameters, and $p\left(\boldsymbol{\theta}\right)$
denotes a prior. A posterior density is proportional to the product
of the likelihood and prior, $p\left(\boldsymbol{\theta}|\mathcal{D}\right)\propto p\left(\mathcal{D}|\boldsymbol{\theta}\right)p\left(\boldsymbol{\theta}\right)$.
Because a posterior is generally intractable, we simulate random draws
from the density using numerical techniques, such as Markov chain
Monte Carlo (MCMC) algorithms. 

A well-known shortcoming of standard Bayesian analysis is that it
needs to fully specify a probabilistic model. Alternative quasi-Bayesian
approaches have been proposed to improve the robustness toward model
misspecification that rely on loss function and moment conditions instead of a probabilistic
model. 

This study considers two types of quasi-Bayesian approaches. In the
first approach, a quasi-likelihood is based on an empirical risk function
$r\left(\boldsymbol{\theta};\mathcal{D}\right)$ (e.g., \citep{Zhang2006,Zhang2006a,Jiang2008,Bissiri2016,Syring2023}).
Hereinafter, we ignore the dependency on $\mathcal{D}$ and write
$r\left(\boldsymbol{\theta}\right)=r\left(\boldsymbol{\theta};\mathcal{D}\right)$.
For instance, consider a linear regression with response variable
$y_{i}$, covariates $\boldsymbol{x}_{i}$, and coefficients $\boldsymbol{\theta}$.
The unknown parameter $\boldsymbol{\theta}$ can be estimated using
an empirical risk function with quadratic losses, $r\left(\boldsymbol{\theta}\right)=N^{-1}\sum_{i=1}^{N}\left(y_{i}-\boldsymbol{\theta}^{\top}\boldsymbol{x}_{i}\right)^{2}$
where $N$ denotes a sample size. Here, a probabilistic assumption
does not have to be imposed on the error term such as the normality;
thus, $\boldsymbol{\theta}$ can be estimated with more robustness.
Given an empirical risk function, a Gibbs posterior is defined as
follows:
\begin{equation}
\pi_{\omega}\left(\boldsymbol{\theta}\right)\propto\left[\exp\left(-Nr\left(\boldsymbol{\theta}\right)\right)\right]^{\omega}p\left(\boldsymbol{\theta}\right).
\end{equation}
In this approach, a scaling parameter (also known as a learning rate
parameter) $\omega\left(>0\right)$ must be selected to calibrate
the spread of the posterior; however, it is not a theoretically or
computationally trivial problem.\footnote{See \citep{Wu2023} for a review of the related literature.}
Recently, Frazier et al. \citep{Frazier2024} proposed a new approach
that specifies a quasi-posterior as follows:
\begin{equation}
\pi\left(\boldsymbol{\theta}\right)\propto\left|\boldsymbol{W}\left(\boldsymbol{\theta}\right)\right|^{-\frac{1}{2}}\exp\left\{ -\frac{N}{2}\bar{\boldsymbol{m}}\left(\boldsymbol{\theta}\right)^{\top}\boldsymbol{W}\left(\boldsymbol{\theta}\right)^{-1}\bar{\boldsymbol{m}}\left(\boldsymbol{\theta}\right)\right\} p\left(\boldsymbol{\theta}\right),\label{eq: quasi-posterior}
\end{equation}
where $\bar{\boldsymbol{m}}\left(\boldsymbol{\theta}\right)=\nabla r\left(\boldsymbol{\theta}\right)$,
$\nabla$ denotes the gradient operator, and $\boldsymbol{W}\left(\boldsymbol{\theta}\right)$
denotes an estimate of the covariance of $\sqrt{N}\bar{\boldsymbol{m}}\left(\boldsymbol{\theta}\right)$.
As argued in \citep{Frazier2024}, this approach can provide proper
uncertainty quantification without selecting $\omega$ under some
regularity conditions. 

The second approach specifies a quasi-likelihood based on a generalized
method of moments (GMM) criterion (e.g., \citep{Kim2002,Chernozhukov2003,Yin2009}).
In a GMM, a statistical problem is formulated under a set of moment
conditions, $E\left[\boldsymbol{m}_{i}\left(\boldsymbol{\theta};\mathcal{D}\right)\right]=\boldsymbol{0}$,
where $\boldsymbol{m}_{i}\left(\cdot\right)$ denotes a vector-valued
function termed a moment function. We shortly denote $\boldsymbol{m}_{i}\left(\boldsymbol{\theta}\right)=\boldsymbol{m}_{i}\left(\boldsymbol{\theta};\mathcal{D}\right)$.
The unknown parameters $\boldsymbol{\theta}$ are estimated by minimizing
a GMM criterion $\bar{\boldsymbol{m}}\left(\boldsymbol{\theta}\right)^{\top}\boldsymbol{W}\left(\boldsymbol{\theta}\right)^{-1}\bar{\boldsymbol{m}}\left(\boldsymbol{\theta}\right)$,
with $\bar{\boldsymbol{m}}\left(\boldsymbol{\theta}\right)=N^{-1}\sum_{i=1}^{N}\boldsymbol{m}_{i}\left(\boldsymbol{\theta}\right)$.
Then, a quasi-posterior is given by 
\begin{equation}
\pi_{\omega}\left(\boldsymbol{\theta}\right)\propto\left[\left|\boldsymbol{W}\left(\boldsymbol{\theta}\right)\right|^{-\frac{1}{2}}\exp\left\{ -\frac{N}{2}\bar{\boldsymbol{m}}\left(\boldsymbol{\theta}\right)^{\top}\boldsymbol{W}\left(\boldsymbol{\theta}\right)^{-1}\bar{\boldsymbol{m}}\left(\boldsymbol{\theta}\right)\right\} \right]^{\omega}p\left(\boldsymbol{\theta}\right).\label{eq: Bayes GMM}
\end{equation}
This approach can be regarded as a special case of (\ref{eq: quasi-posterior})
when the moment conditions are derived under the first-order condition
for an empirical risk function. Although the first approach can only
handle exactly identified models, the second can also handle over-identified
models. For notational convenience, we subsequently focus on (\ref{eq: quasi-posterior}).

The biggest challenge in implementing Bayesian inference with (\ref{eq: quasi-posterior})
is the computational load. The evaluation of $\pi\left(\boldsymbol{\theta}\right)$
is computationally expensive because of the inverse matrix $\boldsymbol{W}\left(\boldsymbol{\theta}\right)^{-1}$
as well as the matrix determinant $\left|\boldsymbol{W}\left(\boldsymbol{\theta}\right)\right|$.
To the best of our knowledge, no studies have addressed this issue.
Yin et al. \citep{Yin2011} proposed a posterior simulator for inference
with (\ref{eq: Bayes GMM}); however, their proposal aimed to address
the numerical stability and can be computationally expensive.

In this study, we propose a new algorithm based on a delayed acceptance
MCMC (DA--MCMC) \citep{Christen2005}, which can be beneficial for
Monte Carlo simulation when the target kernel is expensive to compute.
DA--MCMC can be more computationally efficient than the standard
MCMC because it can shorten the computation of $\boldsymbol{W}\left(\boldsymbol{\theta}\right)^{-1}$.
A single cycle of DA--MCMC has two stages. In the first stage, after
being generated for a new state, the proposal is promoted to the second
stage with probability based on an approximate target density. In
the second stage, the promoted proposal is accepted with probability
based on the exact target density. Through applications to synthetic
and real data problems, we demonstrate that the proposed algorithm
is approximately two times more efficient than the standard MCMC in
terms of the effective sample size per second. 

The remainder of this paper is organized as follows: Section 2 introduces
DA--MCMC, Section 3 compares the proposed algorithm with an existing
algorithm via synthetic and real data examples, and Section 4 concludes
the paper.

\section{Delayed acceptance MCMC}

To reduce computational costs, this study proposes to employ a DA--MCMC
algorithm \citep{Christen2005} which proceeds in two stages. We first
approximate the target posterior distribution $\pi\left(\boldsymbol{\theta}\right)$
with a computationally cheaper surrogate $\pi^{*}\left(\boldsymbol{\theta}\right)$,
which avoids evaluating the most expensive components---specifically,
the inverse and determinant of $\boldsymbol{W}\left(\boldsymbol{\theta}\right)$.
The approximate posterior is constructed by replacing $\boldsymbol{W}\left(\boldsymbol{\theta}\right)$
with the current value $\boldsymbol{W}\left(\boldsymbol{\theta}^{\left(t\right)}\right)$,
as follows:
\[
\pi^{*}\left(\boldsymbol{\theta}\right)=\left|\boldsymbol{W}\left(\boldsymbol{\theta}^{\left(t\right)}\right)\right|^{-\frac{1}{2}}\exp\left\{ -\frac{N}{2}\bar{\boldsymbol{m}}\left(\boldsymbol{\theta}\right)^{\top}\boldsymbol{W}\left(\boldsymbol{\theta}^{\left(t\right)}\right)^{-1}\bar{\boldsymbol{m}}\left(\boldsymbol{\theta}\right)\right\} p\left(\boldsymbol{\theta}\right).
\]

In the first stage, a candidate $\boldsymbol{\theta}^{\prime}$ is
proposed from a proposal distribution $q_{1}\left(\boldsymbol{\theta}^{\prime}|\boldsymbol{\theta}^{\left(t\right)}\right)$.
It is then preliminarily accepted with probability:
\[
\alpha_{1}\left(\boldsymbol{\theta}^{\left(t\right)},\boldsymbol{\theta}^{\prime}\right)=\min\left\{ 1,\;\frac{q_{1}\left(\boldsymbol{\theta}^{\left(t\right)}|\boldsymbol{\theta}^{\prime}\right)\pi^{*}\left(\boldsymbol{\theta}^{\prime}\right)}{q_{1}\left(\boldsymbol{\theta}^{\prime}|\boldsymbol{\theta}^{\left(t\right)}\right)\pi^{*}\left(\boldsymbol{\theta}^{\left(t\right)}\right)}\right\} .
\]
If accepted, the candidate is promoted to the second stage, where
it is evaluated under the full posterior $\pi\left(\boldsymbol{\theta}\right)$.
Otherwise, it is rejected and the chain remains at the current state
$\boldsymbol{\theta}^{\left(t+1\right)}=\boldsymbol{\theta}^{\left(t\right)}$.

In the second stage, the final acceptance probability is computed
as:
\[
\alpha_{2}\left(\boldsymbol{\theta}^{\left(t\right)},\boldsymbol{\theta}^{\prime}\right)=\min\left\{ 1,\;\frac{q_{2}\left(\boldsymbol{\theta}^{\left(t\right)}|\boldsymbol{\theta}^{\prime}\right)\pi\left(\boldsymbol{\theta}^{\prime}\right)}{q_{2}\left(\boldsymbol{\theta}^{\prime}|\boldsymbol{\theta}^{\left(t\right)}\right)\pi\left(\boldsymbol{\theta}^{\left(t\right)}\right)}\right\} ,
\]
where the effective second-stage proposal distribution is corrected
as: 
\[
q_{2}\left(\boldsymbol{\theta}^{\prime}|\boldsymbol{\theta}^{\left(t\right)}\right)=\alpha_{1}\left(\boldsymbol{\theta}^{\left(t\right)},\boldsymbol{\theta}^{\prime}\right)q_{1}\left(\boldsymbol{\theta}^{\prime}|\boldsymbol{\theta}^{\left(t\right)}\right)+\left(1-r\left(\boldsymbol{\theta}^{\left(t\right)}\right)\right)\delta_{\boldsymbol{\theta}^{\left(t\right)}}\left(\boldsymbol{\theta}^{\prime}\right),
\]
with
\[
r\left(\boldsymbol{\theta}^{\left(t\right)}\right)=\int\alpha_{1}\left(\boldsymbol{\theta}^{\left(t\right)},\boldsymbol{\theta}^{\prime}\right)q_{1}\left(\boldsymbol{\theta}^{\prime}|\boldsymbol{\theta}^{\left(t\right)}\right)d\boldsymbol{\theta}^{\prime}.
\]
Here $\delta_{\boldsymbol{\theta}^{\left(t\right)}}\left(\cdot\right)$ denotes the Dirac mass at $\boldsymbol{\theta}^{\left(t\right)}$. 
This structure improves computational efficiency by rejecting implausible
proposals early, while still ensuring exact sampling from the full
posterior distribution.

As argued in \citep{Christen2005}, compared with the standard MCMC,
a single cycle of DA--MCMC is slightly less efficient in terms of
effective sample size per iteration. However, DA--MCMC can be more
efficient in terms of the effective sample size per second because
it does not require the unnecessary computation of $\boldsymbol{W}\left(\boldsymbol{\theta}^{\prime}\right)^{-1}$
and $\left|\boldsymbol{W}\left(\boldsymbol{\theta}^{\prime}\right)\right|$.
As shown in the next section, DA--MCMC is effective in our context.

This study uses an adaptive random walk Metropolis--Hastings kernel
specified as:
\[
q_{1}\left(\boldsymbol{\theta}^{\prime}|\boldsymbol{\theta}^{\left(t\right)}\right)=\mathrm{N}\left(\boldsymbol{\theta}^{\prime}|\boldsymbol{\theta}^{\left(t\right)},\;\varepsilon\boldsymbol{\Sigma}\right),
\]
where $\varepsilon>0$ denotes the step size, $\boldsymbol{\Sigma}$
denotes the $K$-dimensional symmetric positive definite matrix, and
$\mathrm{N}\left(\boldsymbol{x}|\boldsymbol{a},\boldsymbol{B}\right)$
denotes the probability density function of a multivariate normal
distribution with mean $\boldsymbol{a}$ and covariance matrix $\boldsymbol{B}$
evaluated at $\boldsymbol{x}$. The parameters of the proposal kernel
are adaptively selected on the fly of the MCMC algorithm \citep{Haario2001}.
At the $t$-th cycle, $\varepsilon$ is updated as $\log\varepsilon\leftarrow\log\varepsilon+t^{-\varsigma}\left(\bar{\alpha}^{\left(t\right)}-\alpha^{\dagger}\right)$,
where $\bar{\alpha}^{\left(t\right)}$ denotes the average acceptance
probability up to the $t$-th cycle, $\alpha^{\dagger}\in\left(0,1\right)$
denotes the target acceptance probability, and $\varsigma\in\left(0.5,1\right)$
denotes the tuning parameter. In this study, we fix $\varsigma=0.51$.
$\boldsymbol{\Sigma}$ is updated based on the empirical covariance
matrix of $\boldsymbol{\theta}^{\left(1\right)},\boldsymbol{\theta}^{\left(2\right)},...,\boldsymbol{\theta}^{\left(t\right)}$.
The adaptation of $\varepsilon$ and $\boldsymbol{\Sigma}$ is performed
only during the warmup phase and not thereafter.

\section{Application}

\subsection{Synthetic data: Heteroskedastic linear regression}

We considered a heteroskedastic linear regression with synthetic data.
The data-generating process was adopted from \citep{Frazier2024}.
A response variable $y_{i}$ was generated from a normal distribution,
$y_{i}\sim\mathcal{N}\left(\boldsymbol{\theta}^{\top}\boldsymbol{x}_{i},\sigma_{i}^{2}\right)$,
where a vector of covariates $\boldsymbol{x}_{i}=\left(x_{1,i},x_{2,i},...,x_{K,i}\right)^{\top}$
with $x_{1,i}=1$, $\boldsymbol{\theta}\in\mathbb{R}^{K}$ denotes
a vector of coefficients, and the variance is specified as $\sigma_{i}^{2}=\left(1+\left|x_{2,i}\right|^{2}+\left|x_{3,i}\right|^{2}\right)/3$.
The true values were set to $\boldsymbol{\theta}=\left(1,1,1,0,...,0\right)^{\top}$.
The covariates except $x_{1,i}$ were simulated from the independent
standard normal distribution. The unknown parameters were estimated
via an empirical risk function with quadratic losses, $r\left(\boldsymbol{\theta}\right)=N^{-1}\sum_{i=1}^{N}\left(y_{i}-\boldsymbol{\theta}^{\top}\boldsymbol{x}_{i}\right)^{2}$.
The prior of $\boldsymbol{\theta}$ was specified as $\boldsymbol{\theta}\sim\mathcal{N}\left(\boldsymbol{0}_{K},100^{2}\boldsymbol{I}_{K}\right)$. 

We considered four settings with combinations of $N\in\left\{ 100,1000\right\} $
and $K\in\left\{ 5,20\right\} $. In each posterior simulation, a
total of 20,000 draws were generated and the last 10,000 draws were
used for evaluation. The numerical performance was evaluated based
on multivariate effective sample size (multiESS) \citep{Vats2019}.
\footnote{The programs were executed on MATLAB (R2024b) on an Ubuntu desktop
(22.04.5 LTS) running on an AMD Ryzen Threadripper 3990X 2.9 GHz 64-core
processor.}

The choice of the target acceptance rate $\alpha^{\dagger}$ is of
practical importance. However, no specific theoretical guidance on
this matter exists, as the theoretical properties of DA--MCMC algorithms
have not been investigated thoroughly. To address this gap, we empirically
examine the choice of $\alpha^{\dagger}$ using a simulation study.
Figure 1 displays the medians of multiESS/iter for different target
acceptance rates $\alpha^{\dagger}\in\left\{ 0.02,0.04,...,0.4\right\} $
based on 1,000 independent runs for each setting. Except for the case
with $N=100$ and $K=20$, multiESS/iter is maximized approximately
0.2 to 0.3. Thus, adapting the DA--MCMC with a random walk Metropolis--Hastings
steps to achieve an acceptance rate of 25\% ($\alpha^{\dagger}=0.25$)
can be recommended as a default option. However, selecting $\alpha^{\dagger}=0.25$
can be suboptimal in certain cases, such as when $N$ is insufficiently
large relative to $K$. As an experiment, we set $\alpha^{\dagger}=0.05$
and ran DA--MCMC on the synthetic data with $N=100$ and $K=20$
to obtain an increased multiESS/s of 1,515 (compared with 993 having
$\alpha^{\dagger}=0.25$). Further investigation into the optimal
selection of $\alpha^{\dagger}$ will be the focus of future research. 

\begin{figure}
\caption{MultiESS/iter for different target acceptance rate}

\medskip{}

\begin{centering}
\includegraphics[scale=0.4]{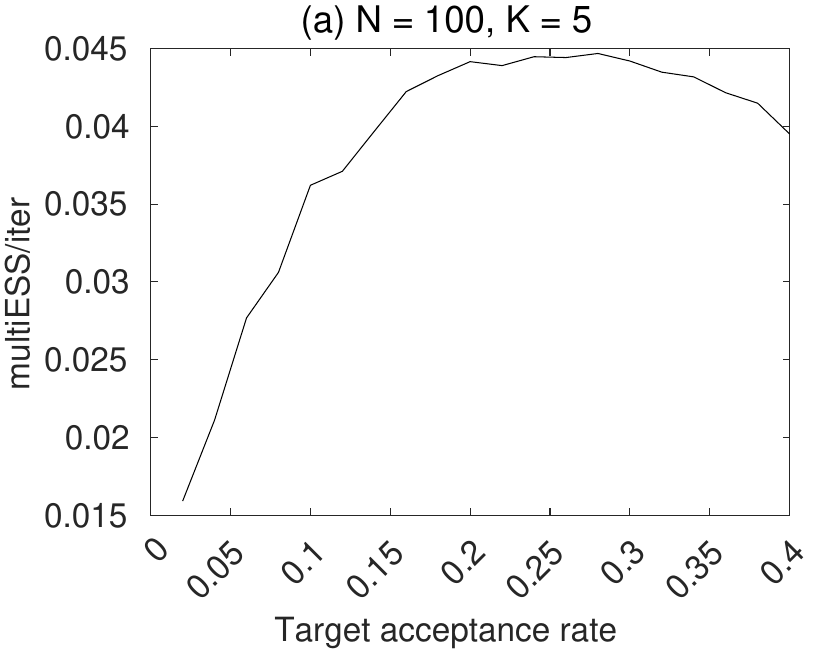}\quad{}\includegraphics[scale=0.4]{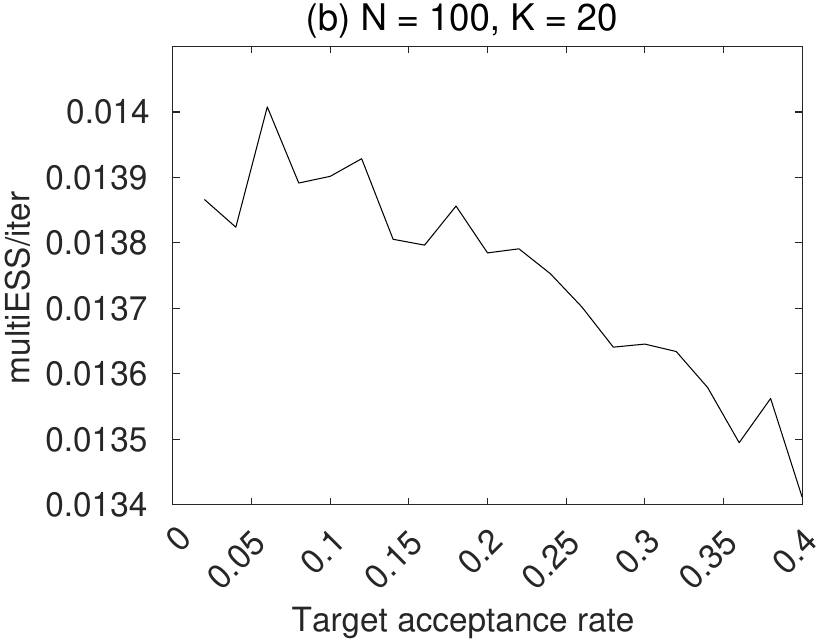}
\par\end{centering}
\medskip{}

\centering{}\includegraphics[scale=0.4]{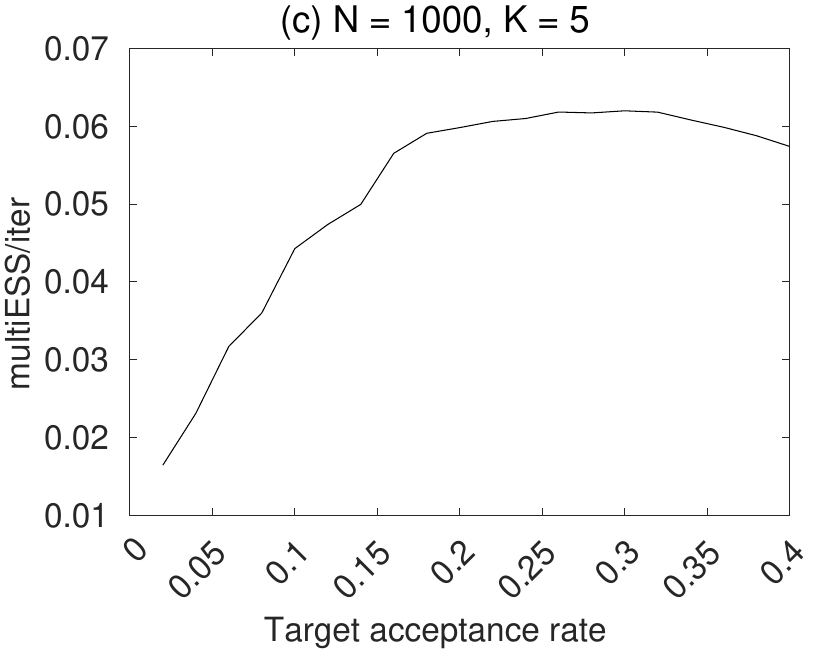}\quad{}\includegraphics[scale=0.4]{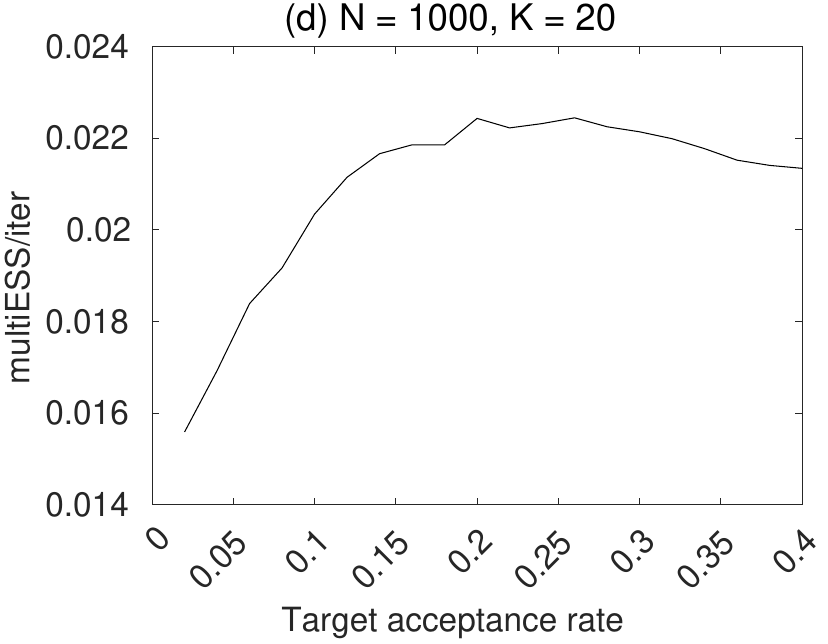}
\end{figure}

Tables 1 and 2 list the medians of the multiESS per iteration (multiESS/iter)
and multiESS per second (multiESS/s) for 1,000 runs, respectively.
As expected, the multiESS/iter for DA--MCMC tended to be slightly
smaller than that for the standard MCMC; however, the difference was
small. In contrast, the multiESS/s for DA--MCMC was approximately
1.5 to 2 times larger than that for the standard MCMC. As shown in
Table 1, the absolute level of efficiency per iteration for this type
of algorithm is low, requiring the generation of long chains to converge.
Therefore, improving efficiency per unit time is an important contribution.

\begin{table}
\caption{Multivariate effective sample size per iteration}

\medskip{}

\centering{}%
\begin{tabular}{lrrcrr}
\hline 
\multirow{2}{*}{Model} & \multirow{2}{*}{$N$} & \multirow{2}{*}{$K$} &  & \multicolumn{2}{c}{multiESS/iter}\tabularnewline
\cline{5-6}
 &  &  &  & MCMC & DA--MCMC\tabularnewline
\hline 
\multirow{4}{*}{Linear Reg.} & \multirow{2}{*}{100} & 5 &  & 0.053 & 0.045\tabularnewline
\cline{3-6}
 &  & \multirow{1}{*}{20} &  & 0.014 & 0.014\tabularnewline
\cline{2-6}
 & \multirow{2}{*}{1000} & 5 &  & 0.062 & 0.062\tabularnewline
 &  & \multirow{1}{*}{20} &  & 0.023 & 0.022\tabularnewline
\hline 
IV Reg. & 64 & 6 &  & 0.020 & 0.021\tabularnewline
\hline 
\end{tabular}
\end{table}

\begin{table}
\caption{Multivariate effective sample size per second}

\medskip{}

\centering{}%
\begin{tabular}{lrrrrr}
\hline 
\multirow{2}{*}{Model} & \multirow{2}{*}{$N$} & \multirow{2}{*}{$K$} &  & \multicolumn{2}{c}{multiESS/s}\tabularnewline
\cline{5-6}
 &  &  &  & MCMC & DA--MCMC\tabularnewline
\hline 
\multirow{4}{*}{Linear Reg.} & \multirow{2}{*}{100} & 5 &  & 4,064 & 6,449\tabularnewline
 &  & \multirow{1}{*}{20} &  & 467 & 993\tabularnewline
\cline{2-6}
 & \multirow{2}{*}{1000} & 5 &  & 2,494 & 4,834\tabularnewline
 &  & 20 &  & 204 & 515\tabularnewline
\cline{2-6}
IV Reg. & 64 & 6 &  & 160 & 293\tabularnewline
\hline 
\end{tabular}
\end{table}

To further investigate the behavior of DA--MCMC, Figure 2 presents
the relative frequency of the second-stage acceptance probability
for linear regression using synthetic data. Table 3 summarizes the
results of the 25th, 50th, and 75th percentiles of these probabilities.
Except for the case with $N=100$ and $K=20$, most of the observed
second-stage acceptance probabilities were close to 1. This indicates
that once a proposal passes the first-stage screening, it is almost
certain that it will be considered acceptable in the second stage.
However, in the most challenging task, where $N=100$ and $K=20$
(where $K$ is large relative to $N$), the second-stage acceptance
probabilities were widely dispersed and ranged from 0 to 1. The median
second-stage acceptance probability was 0.250, suggesting that most
of the promoted proposals were rejected. This implies that the first-stage
posterior was not a good approximation, because the contributions
of $\boldsymbol{W}\left(\boldsymbol{\theta}\right)^{-1}$ and $\left|\boldsymbol{W}\left(\boldsymbol{\theta}\right)\right|$
to the (true) posterior were likely significant.

\begin{figure}
\caption{Relative frequency of second-stage acceptance probability}

\medskip{}

\begin{centering}
\includegraphics[scale=0.4]{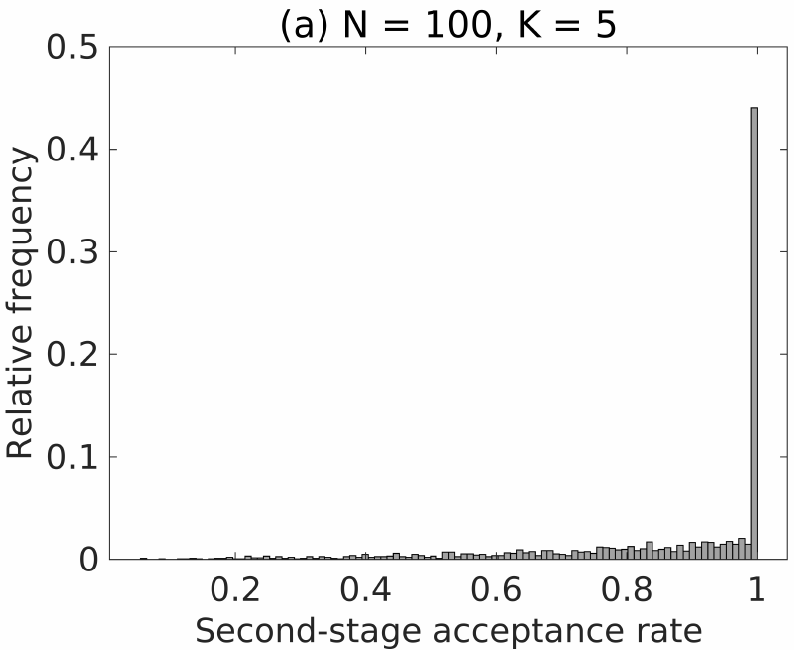}\quad{}\includegraphics[scale=0.4]{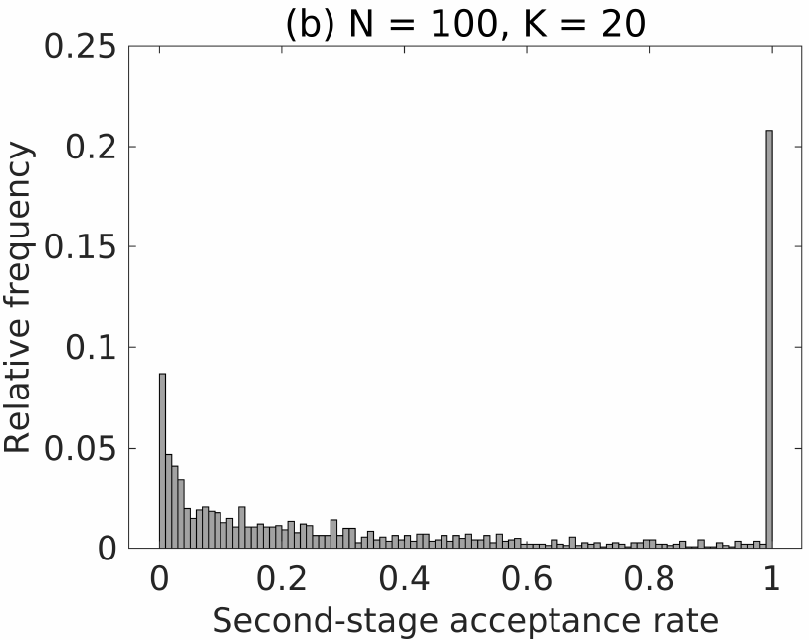}
\par\end{centering}
\medskip{}

\centering{}\includegraphics[scale=0.4]{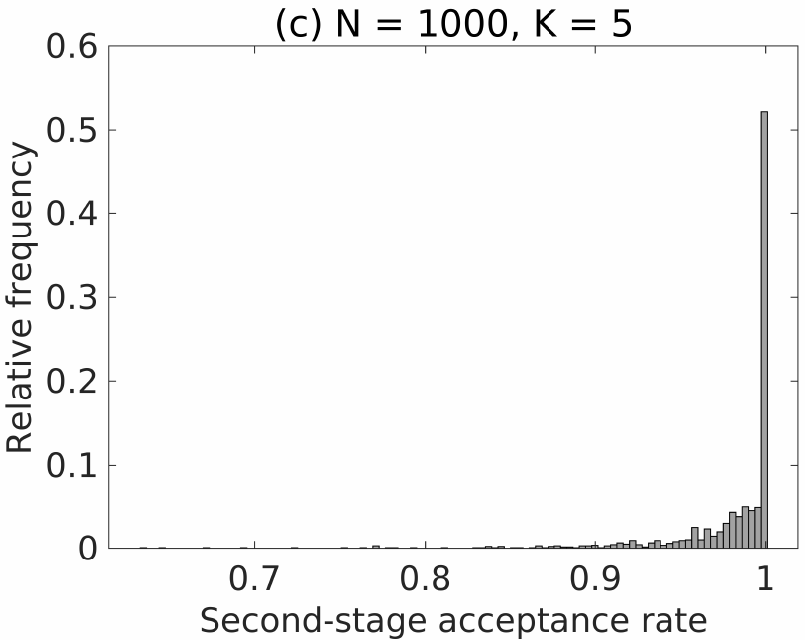}\quad{}\includegraphics[scale=0.4]{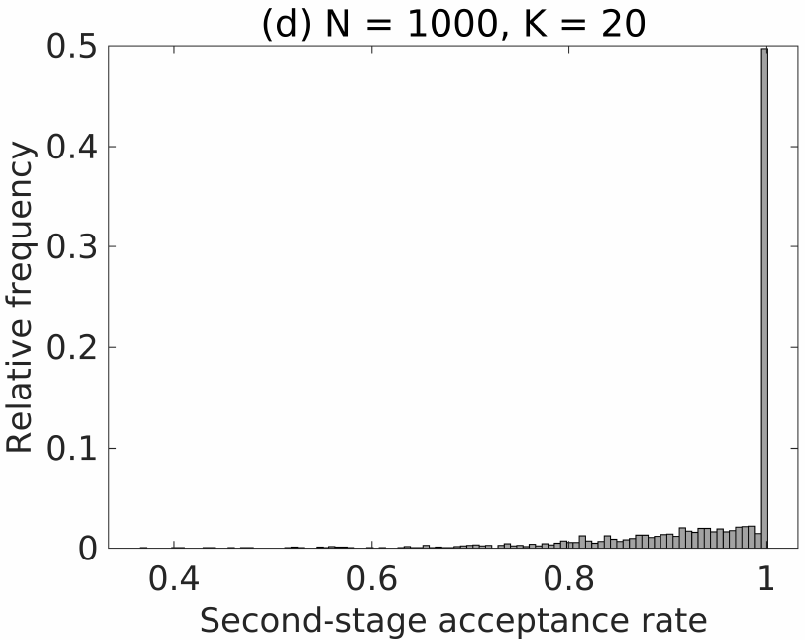}
\end{figure}

\begin{table}
\caption{Second-stage acceptance probability}

\medskip{}

\centering{}%
\begin{tabular}{rrrrr}
\hline 
\multirow{2}{*}{$N$} & \multirow{2}{*}{$K$} & \multicolumn{3}{c}{Percentile}\tabularnewline
\cline{3-5}
 &  & $P_{25}$ & $P_{50}$ & $P_{75}$\tabularnewline
\hline 
\multirow{2}{*}{100} & 5 & 0.760 & 0.958 & 1.000\tabularnewline
 & 20 & 0.062 & 0.250 & 0.798\tabularnewline
\hline 
\multirow{2}{*}{1000} & 5 & 0.978 & 0.999 & 1.000\tabularnewline
 & 20 & 0.907 & 0.992 & 1.000\tabularnewline
\hline 
\end{tabular}
\end{table}

\subsection{Real data: Instrumental variable regression}

We estimated an instrumental variable regression using a real dataset.
The dataset is obtained from Acemoglu et al. \citep{Acemoglu2001,Acemoglu2012}.\footnote{The dataset is downloaded from the OPENICPSR website.

https://www.openicpsr.org/openicpsr/project/112564/version/V1/view} They investigated the effect of the expropriation risk ($x_{i}$)
on the gross domestic product per capita (in logs) ($y_{i}$). They
used the European settler mortality (in logs) as an instrument ($z_{i}$)
to address endogeneity. The vector of controls $\boldsymbol{w}_{i}$
includes a constant, latitude, and continents dummies, namely, dummies
for Africa, Asia, and the former British colonies (Australia, Canada,
New Zealand, and the United States). 

The model is expressed as follows:
\begin{eqnarray*}
x_{i} & = & g\left(z_{i},\boldsymbol{w}_{i}\right)+v_{i},\\
y_{i} & = & \mu+\beta x_{i}+\boldsymbol{\gamma}^{\top}\boldsymbol{w}_{i}+u_{i},
\end{eqnarray*}
where $v_{i}$, and $u_{i}$ denote error terms. Here, $g\left(\cdot\right)$
denotes an unknown function. The unknown parameters $\boldsymbol{\theta}=\left(\mu,\beta,\boldsymbol{\gamma}^{\top}\right)^{\top}$
are inferred under the following moment conditions:
\[
\bar{\boldsymbol{m}}\left(\boldsymbol{\theta}\right)=\frac{1}{N}\sum_{i=1}^{N}\left(y_{i}-\left(\mu+\beta x_{i}+\boldsymbol{\gamma}^{\top}\boldsymbol{w}_{i}\right)\right)\left(\begin{array}{c}
z_{i}\\
\boldsymbol{w}_{i}
\end{array}\right).
\]
For $\boldsymbol{\theta}$, we use a normal prior, $\boldsymbol{\theta}\sim\mathcal{N}\left(\boldsymbol{0}_{K},100^{2}\boldsymbol{I}_{K}\right)$.
This approach has notable advantages over standard Bayesian approaches
to instrumental variable regression, such as \citep{Lopes2014,Wiesenfarth2014,Spanbauer2024}.
First, it does not require the estimation of $g\left(\cdot\right)$.
Second, it does not specify the distributions of $v_{i}$ and $u_{i}$.
The sample size is $N=64$ and the number of unknown parameters is
$K=6$. We discarded the first 100,000 draws and used the last 1,000,000
draws for analysis.

The comparison of the numerical performance based on 1,000 runs is
summarized in the last rows of Tables 1 and 2. As in the application
to synthetic data above, although the multiESS/iter for DA--MCMC
was virtually indistinct from that for the standard MCMC, the multiESS/s
for DA--MCMC was much larger than that for the standard MCMC. Figure
3 compares the simulated posterior distributions of $\beta$ using
the two algorithms. The difference between the two was virtually negligible.
The result shows that these two algorithms can generate random numbers
from the same distribution, despite their significantly different
computational efficiencies.

\begin{figure}
\caption{Posterior distribution of $\beta$}

\medskip{}

\centering{}\includegraphics[scale=0.4]{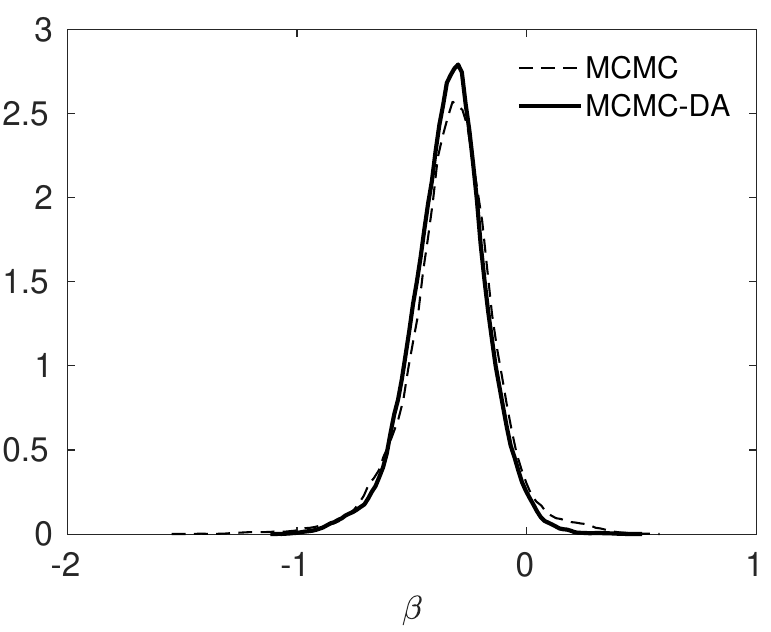}
\end{figure}

\section{Conclusion}

This study introduced a computationally efficient algorithm for Bayesian
inference, specifically tailored for quasi-Bayesian approaches using
DA--MCMC. By leveraging the two-stage screening and promotion process
of DA--MCMC, we demonstrated that it significantly reduced the computational
burden of evaluating the quasi-posterior, particularly when considering
high-dimensional models and costly matrix computations. Through synthetic
and real data applications, we demonstrated that although the DA--MCMC
algorithm slightly decreased effective sample size per iteration compared
with the standard MCMC, it offered a substantial improvement in terms
of computational efficiency, approximately doubling the effective
sample size per second. This efficiency gain holds promise for broader
applications in complex statistical models where posterior simulation
can otherwise be prohibitively slow. Future work will explore optimizing
the algorithm further for large-scale problems or its application
in other quasi-Bayesian frameworks.

\section*{Acknowledgements}

This work was supported by JSPS KAKENHI Grant Numbers 20K22096 and
25K21168.

\bibliographystyle{spmpsci}
\bibliography{reference}

\end{document}